\documentclass[11pt]{article}
\pdfoutput=1
\usepackage{graphicx}
\usepackage{epstopdf} 
\usepackage{epsfig}
\usepackage{amsmath}
\usepackage{amssymb}
  \usepackage{color}
  \usepackage{verbatim}
\usepackage{multirow}
\usepackage[bottom]{footmisc}
\usepackage{cite,url}
\usepackage{comment}
\usepackage{subfig}
\usepackage{textcomp}
\usepackage{booktabs,multirow,tabularx}

\newcommand\php{\phi^+}
\newcommand\phm{\phi^-}
\newcommand\phu{h}
\newcommand\phd{\phi_2}

\newcommand\Phid{\Phi^\dagger}

\newcommand{\nn}{\nonumber}

\newcommand\sss{\scriptscriptstyle}

\newcommand{\ord}{\mathcal{O}}
\newcommand{\tril}{\lambda_{3}}
\newcommand{\trilsm}{\tril^{\rm SM}}
\newcommand{\qual}{\lambda_{4}}
\newcommand{\mh}{m_{ \sss H}}
\newcommand{\Mh}{M_{ \sss H}}
\newcommand{\mw}{m_{ \sss W}}
\newcommand{\mz}{m_{ \sss Z}}
\newcommand{\mt}{m_{t}}

\newcommand{\ktre}{\kappa_{\lambda}}

\def\beq{\begin{equation}}
\def\beqn{\begin{eqnarray}}
\def\eeq{\end{equation}}
\def\eeqn{\end{eqnarray}}
\def\beal{\begin{align}}
\def\endal{\end{align}}

\long\def\symbolfootnote[#1]#2{\begingroup%
\def\thefootnote{\fnsymbol{footnote}}\footnote[#1]{#2}\endgroup}

\begin{document}\hfill
\vspace*{1cm}
\color{black}
\begin{center}
{\Large \bf  \color{blue} The effect of an  anomalous  Higgs trilinear 
 self-coupling on the $h \to \gamma \, Z$ decay}

\bigskip\color{black}\vspace{.4cm}
{\large\bf G.~Degrassi, M.~Vitti}
\\[7mm]
{\it  Dipartimento di Matematica e Fisica, Universit{\`a} di Roma Tre and \\
 INFN, sezione di Roma Tre, I-00146 Rome, Italy}\\[1mm]
\end{center}
\symbolfootnote[0]{{\tt e-mail:}}
\symbolfootnote[0]{{\tt degrassi@fis.uniroma3.it}}
\symbolfootnote[0]{{\tt marco.vitti@uniroma3.it}}
\bigskip
\vspace{.5cm}

\centerline{\large\bf Abstract}
\begin{quote}
  We compute the two-loop effects induced by an anomalous Higgs
  trilinear self-coupling in the partial decay width $h \to \gamma Z$. The
  computation is performed using the anomalous coupling approach
  working in the unitary gauge and in a theory in which the anomalous
  coupling is generated via the addition to the scalar
  potential part of the Standard Model Lagrangian of an (in)finite tower of
  $(\Phid \Phi)^n$ terms. The former
  computation is automatically finite while the latter requires the
  renormalization of the lowest order contribution.  We discuss the
  renormalization conditions that should be employed in order to
  obtain the same result in the two approaches. We find that the
  $h \to \gamma Z$ process is one of the most sensitive mode to an
  anomalous trilinear Higgs  self-coupling.  As a by-product of
  this work we confirm one of  two different results present in
  the literature concerning the contribution of an anomalous Higgs
  trilinear coupling in the $h \to \gamma \gamma$ decay.
\end{quote}
\thispagestyle{empty}
\newpage
\section{Introduction}
The properties of the scalar particle with mass around 125 GeV discovered
at the Large Hadron Collider (LHC) in 2012
\cite{Chatrchyan:2012xdj,Aad:2012tfa} have been extensively studied since
its observation. These studies show strong evidence that the coupling of this
particle to fermions and vector bosons are compatible within $10-20 \%$ with
those of the Higgs boson as predicted in the Standard Model (SM) of elementary
particles.

The complete identification of the scalar particle discovered in 2012 with
the Higgs boson of the SM requires also the  study of the Higgs self
interactions that  come from  the scalar potential
part in the SM Lagrangian. In  the SM, the  Higgs potential in the unitary
gauge reads
\beq
V(\phu) =  \frac{\mh^2}{2} \phu^2 + \tril v  \phu^3 +
\frac{\lambda_4}4 \phu^4  \label{eq:potun}
\eeq
where the Higgs mass ($\mh$) and the trilinear $(\tril)$ and quartic $(\qual)$ 
interactions are linked by the relations 
$\qual^{\rm SM}=\trilsm = \lambda =\mh^2/(2\,v^2)$,
where $v=(\sqrt2 \,G_\mu)^{-1/2}$ is the vacuum expectation value, and 
$\lambda$ is the coefficient of the $(\Phid \Phi)^2$
interaction, $\Phi$ being the Higgs doublet field.

The experimental verification of these relations relies on the  measurements
of double-Higgs and triple-Higgs productions. However, since the cross sections
for these processes are quite small, constraining $\tril$ and $\lambda_4$
couplings within few times their predicted SM value is already extremely
challenging. In the case of double Higgs production at present only
exclusion limits are available. The most stringent result, coming from
the ATLAS combination of the $b\bar{b}b\bar{b},\:  b\bar{b}\tau \tau$ and
$b\bar{b} \gamma \gamma$ channels, allows to set a bound $-5\, \trilsm <
\tril < 12\, \trilsm$ at 95\% C.L. \cite{Aad:2019uzh}. It is not yet clear if
double Higgs
production will be observed at the end of the high luminosity (HL) period
of LHC with a collected luminosity of 3000 fb$^{-1}$ or just $\cal{O}$ $(1)$
bounds on $\tril$ are going to be set. Concerning $\qual$,
given the smallness of the  triple-Higgs production cross section
(around $0.1$ fb at $\sqrt{s}= 14$ TeV)  $\qual$ will be only very loosely
constrained at  the HL-LHC.

In order to  constrain the trilinear Higgs self-coupling,
a complementary strategy based  on  precise measurements was proposed.
In this approach the effects induced at the loop level on
various processes by a modified $\tril$ coupling are studied.
This approach builds on the assumption that New Physics (NP) couples to
the SM via the Higgs potential in such a way that the lowest-order Higgs
couplings to the other fields of the SM (and in particular to the top quark and
vector bosons) are still given by the SM prescriptions or,
equivalently, modifications to these couplings are so small that do
not swamp the loop effects one is considering.
This strategy was first applied to $ZH$ production at an $e^+ e^-$ collider in
Ref.~\cite{McCullough:2013rea}, later to Higgs production and decay modes
at the LHC \cite{Degrassi:2016wml,Gorbahn:2016uoy,Bizon:2016wgr,
  Maltoni:2017ims,Maltoni:2018ttu,Borowka:2018pxx,Gorbahn:2019lwq} and also to
the study of
electroweak precision observables \cite{Degrassi:2017ucl,Kribs:2017znd}.
Using this strategy a recent analysis of the ATLAS Collaboration
set a more stringent bound, $-2.3\, \trilsm < \tril < 10.3\, \trilsm$
at 95\% C.L.,
by combining the single Higgs boson analyses targeting the
$\gamma \gamma,\, Z Z^\star,\, W W^\star,\, \tau^{+} \tau^- $  and
$b \overline{b}$
decay channels and the double Higgs boson analyses in the
$b\bar{b}b\bar{b},\:  b\bar{b}\tau \tau$ and
$b\bar{b} \gamma \gamma$ decay channels using data collected at
$\sqrt{s} = 13$ TeV \cite{ATLAS:2019pbo}. 

The aim of this  work is twofold. On one side we continue the program of
identifying in single Higgs processes the $\tril$-dependent loop contributions
by examining the $\phu \to \gamma Z$ decay. 
On the other side we use this decay to show  in detail,
for contributions that arise at two-loop level, the equivalence,
at the loop order we are working, of two approaches. In the first one,
following Ref.~\cite{Degrassi:2016wml}, the $\tril$-dependent loop
contributions are studied via the introduction of a rescaling
factor multiplying the trilinear SM coupling, $\tril = \ktre \trilsm$, with
$\trilsm\equiv G_\mu \,\mh^2/\sqrt{2} $,
working in the unitary gauge (UG): this is the so called $\kappa$-framework or
anomalous coupling approach. The second one is based on the modification of
the SM scalar potential via the addition of higher-dimensional operators that
affect only the Higgs self-coupling, i.e an (in)finite tower of
$(\Phid \Phi)^n$ terms $(n >2)$. 
  The latter, when only the term $n=3$ is
assumed, corresponds to the SM effective field theory (SMEFT) approach with
one single dimension-six operator, the $O_6 = (\Phid \Phi)^3$ one.

The calculation of $\tril$-dependent loop contributions
in the  $\phu \to \gamma Z$ decay shares many similarities with the analogous
calculation for $\phu \to \gamma \gamma$. For this process the results
in the literature obtained using the anomalous coupling approach
\cite{Degrassi:2016wml} and the SMEFT approach \cite{Gorbahn:2016uoy}
are not in agreement. As by-product of our  $\phu \to \gamma Z$ analysis we
resolve this discrepancy.

The paper is organized as follows. In section \ref{sec:2} we present
the general structure of the $\tril$-dependent contribution in
$\Gamma(\phu \to \gamma Z)$ outlining the (approximate) way this
contribution is evaluated. In section \ref{sec:3} we discuss the
renormalization conditions that should be imposed on the scalar
potential of a $(\Phid \Phi)^n$ theory in order to obtain a result for the
anomalous trilinear contribution that is identical to the one obtained
in the $\kappa$-framework working in the UG where the renormalization
of the one-loop contribution is not needed. 
Section \ref{sec:4} presents our results for  the $\tril$-dependent
contribution in the partial width $\Gamma(\phu \to \gamma Z)$ and its
corresponding branching ratio (BR). Finally, in the Conclusions we shortly
discuss the  contribution due to an anomalous  trilinear Higgs coupling
in $\Gamma(\phu \to \gamma \gamma)$ confirming the equivalence between
the $\kappa$-framework and the SMEFT with the $O_6$ operator.

\section{$\tril$-dependent contribution in $\Gamma (h \to \gamma Z)$}
\label{sec:2}
We begin by recalling the structure of the $\tril$-dependent contribution in single
Higgs processes as discussed in Ref.\cite{Degrassi:2016wml}. This contribution
arises from next-to-leading order (NLO) electroweak (EW) corrections and can
be organized in two categories: a universal part proportional to $(\ktre)^2$
due to the wave-function renormalization of the external Higgs boson, and
a process-dependent part linear in $\ktre$ that in general depends on the
kinematics of the process under consideration.

Specializing to the   $h \to \gamma Z$ decay we write for the
$\tril$-dependent contribution in the width
\beq
\Gamma_{\tril} = Z_H\,\Gamma_{\rm{LO}} \left( 1 + \ktre C_1 \right)\,,
\label{NLOEWSM}
\eeq   
with
\beqn
Z_H &=& \frac1{1- \ktre^2\, \delta Z_H}, \\
\delta Z_{H} &=& -\frac{9}{16} \, 
\frac{G_\mu\,\mh^2 }{\sqrt{2}\, \pi^2} \left(\frac{2\pi}{3\sqrt{3}}-1\right)=
-1.536 \cdot 10^{-3} \,.
\label{deltaZH}
\eeqn
In eq.(\ref{NLOEWSM}) $C_1$ is the process-dependent contribution that will
be presented in this paper while $\Gamma_{\rm{LO}}$ stands for the LO prediction.
Neglecting $\ord(\ktre^3\,\alpha^2)$ the relative corrections induced by an
anomalous trilinear Higgs  self-coupling can be expressed as 
\beq
d \Gamma_{\tril} \equiv 
\frac{\Gamma_{\rm{NLO}} - \Gamma^{\rm SM}_{\rm{NLO}}}{\Gamma_{\rm{LO}}} =
(\ktre-1) C_1+(\ktre^2-1) C_2 \,,
\label{corr}
\eeq
where $\Gamma^{\rm SM}_{\rm{NLO}}$ is the NLO SM result\footnote{In the SM NLO
  result only the  contribution proportional to $\trilsm$ is included.}
and
\beq
C_2 = \frac{\delta Z_H}{(1- \ktre^2 \delta Z_H)}~.
\label{eqC2}
\eeq
The range of validity of eq.(\ref{corr}) can be identified according to
Ref.\cite{Degrassi:2016wml} with $|\ktre| \lesssim 20$ where this bound was 
derived via an estimate of the missing terms in the perturbative calculation
of single Higgs processes.

Because the photon does not couple directly to neutral particles,
the decay process $h \rightarrow \gamma Z$ receives contribution at LO from
one-loop diagrams. Then, the evaluation of the $C_1$ coefficient requires
a two-loop calculation. An exact evaluation of the $\tril$-dependent two-loop
diagrams is not possible. Then, we employ the same strategy used in the
analogous calculation of the $\Gamma( h \to \gamma \gamma)$
\cite{Degrassi:2016wml}, namely the relevant diagrams are computed via a
Taylor expansion in the external momenta.
Calling $q_1$ and $q_2$ the momenta of the photon and of the $Z$, respectively,
we make an expansion in the parameters $q^2_1/(4 m^2)$, $q^2_2/(4 m^2)$
and $(q_1 \cdot q_2)/(4 m^2)$  where $m$ is the mass of any  particle
running into
the loops, i.e. $m=\mt, \,\mh,\, \mw,\,\mz$, and at the end of the computation
we set $q_1^2 =0 , \,q^2_2= \mz^2$ and  $(q_1 \cdot q_2)= (\mh^2 - \mz^2)/2$.
We point out that our expansion
parameters are all smaller than 1 although not always very small.

In order to test the consistency of our small-momentum expansion
approach we first compare  the LO contribution computed exactly, 
with its evaluation via a small momentum expansion in the UG. 

The $h \to \gamma Z$ decay width can be written,
\begin{equation}
\Gamma(h \to \gamma Z) = 
\frac{G_\mu^2 \mz^2  \alpha ~ \mh^3}{64 \pi^4} \Biggl(1-\frac{\mz^2}{\mh^2}
\Biggr)^3 |\mathcal{F} |^2
\end{equation}
with 
\begin{equation}
\mathcal{F} = N_c  Q_t ~ (I_{3 t}-2Q_t \sin^2 \theta_W) \mathcal{F}_t + \cos^2 \theta_W \mathcal{F}_W
\end{equation}
where $\theta_W$ is the weak angle, while $\mathcal{F}_t$ and $\mathcal{F}_W$
are the fermionic and bosonic contributions to the  amplitude, respectively.
In the former we are
considering only the dominant top quark contribution
that sets the color factor, the electric charge and the weak isospin to
be $N_c=3, \,Q_t =2/3,\, I_{3t} =1/2$.

At LO the $\mathcal{F}_t$ and $\mathcal{F}_W$ terms can be written
\cite{Cahn:1978nz,Bergstrom:1985hp}

\begin{equation}
\mathcal{F}_t^{\rm{LO}} = 2  \Bigl[I_1(1/h_{4t}, 1/z_{4t}) - I_2 (1/h_{4t}, 1/z_{4t}) \Bigr]
\label{eq:Ft1l_analytic}
\end{equation}

\beqn
\mathcal{F}_W^{\rm{LO}} &=&   4(3 - \tan^2 \theta_W) I_2(1/h_{4w}, 1/z_{4w})\nn  \\
&+&  \Bigl[ (1+ 2 h_{4w} ) \tan^2 \theta_W - (5+2 h_{4w} ) \Bigr]
I_1 (1/h_{4w}, 1/z_{4w}) 
\label{eq:FW1l_analytic}
\eeqn
where $h_{4 i} = \mh^2/(4 m_i^2)$ and $z_{4 i} = \mz^2/(4 m_i^2)$. 
The functions $I_1$ and $I_2$ are defined as

\beqn
I_1 (\tau, \lambda) &=& \frac{\tau \lambda}{2(\tau -\lambda)} + \frac{\tau^2 \lambda^2}{2(\tau -\lambda)^2} \Bigl[f(\tau) -f(\lambda) \Bigr] + \frac{\tau^2 \lambda}{(\tau - \lambda)^2} \Bigl[ g(\tau) - g(\lambda) \Bigr], \nn \\
&& \ \label{Iuno}\\
I_2 (\tau, \lambda) &=& - \frac{\tau \lambda}{2(\tau -\lambda)} \Bigl[f(\tau)-
  f(\lambda) \Bigr],
\eeqn
where  (for $x \geq 1$)
\begin{equation}
f(x) = \arcsin^2 \Biggl(\frac{1}{\sqrt{x}} \Biggr)\, , \qquad g(x) = \sqrt{x -1} \arcsin \Biggl(\frac{1}{\sqrt{x}} \Biggr) .
\end{equation}

Using the small-momentum expansion we obtain the following expressions
for the fermionic and bosonic contributions
\beqn
  \mathcal{F}_t^{\rm{LO}} &=&   \frac{2}{3}+\frac{1}{45}\, h_{4t} (7+11 z_h) +
  \frac{4}{315} \,h_{4t}^2 \left(5+8 z_h+11 z_h^2\right)\nn  \\
&+&  \frac{4}{1575}\, h_{4t}^3  \left(13+21 z_h+29 z_h^2+37 z_h^3 \right) \nn \\
 &+&  \frac{128}{51975}\, h_{4t}^4
  \left(8+13 z_h+18 z_h^2+23 z_h^3+28 z_h^4 \right)  + \mathcal{O}(h_{4t}^5)
  \label{Ftexp}\\
\mathcal{F}_W^{\rm{LO}} &= & 
- 7-\frac{2}{15}\, h_{4w} (11-37 z_h)+
\frac{4}{105}\, h_{4w}^2 \left(-19-z_h+31 z_h^2\right)  \nn \\
&-& \frac{8}{1575}\, h_{4w}^3 \left(87+19 z_h-39 z_h^2-97 z_h^3\right) \nn \\
   &-& \frac{16}{17325}\, h_{4w}^4
\left(328+93 z_h-32 z_h^2-157 z_h^3-282 z_h^4\right) + \mathcal{O}(h_{4w}^5)
\label{Fwexp}
\eeqn
where $z_h= \mz^2/\mh^2$.

We checked that eq.(\ref{Ftexp}) and eq.(\ref{Fwexp}) match exactly the
expansion  of the expressions in eq.(\ref{eq:Ft1l_analytic}) and
 eq.(\ref{eq:FW1l_analytic}) in the limit of small $h_{4 i}$ and $z_{4 i}$.
 
 Neglecting the last known contribution in eqs.(\ref{Ftexp},\ref{Fwexp}),
 i.e.  $\mathcal{O}(h_{4t}^4)$ and $\mathcal{O}(h_{4w}^4)$, the
numerical result for the decay width obtained with the small-momentum
expansion differs from the evaluation of the exact expression by
2.6\%. Including also this last term the difference reduces to 1.3\%.
Then, we  can  estimate that the evaluation of the two-loop contribution via a
small momentum expansion including $\mathcal{O}(h_{4t}^3)$ and
$\mathcal{O}(h_{4w}^3)$ is expected to differ from the exact result by
 $\mathcal{O}(5\%)$.

\begin{figure}[t]
\centering
	\subfloat[]{
	\includegraphics[scale=0.20]{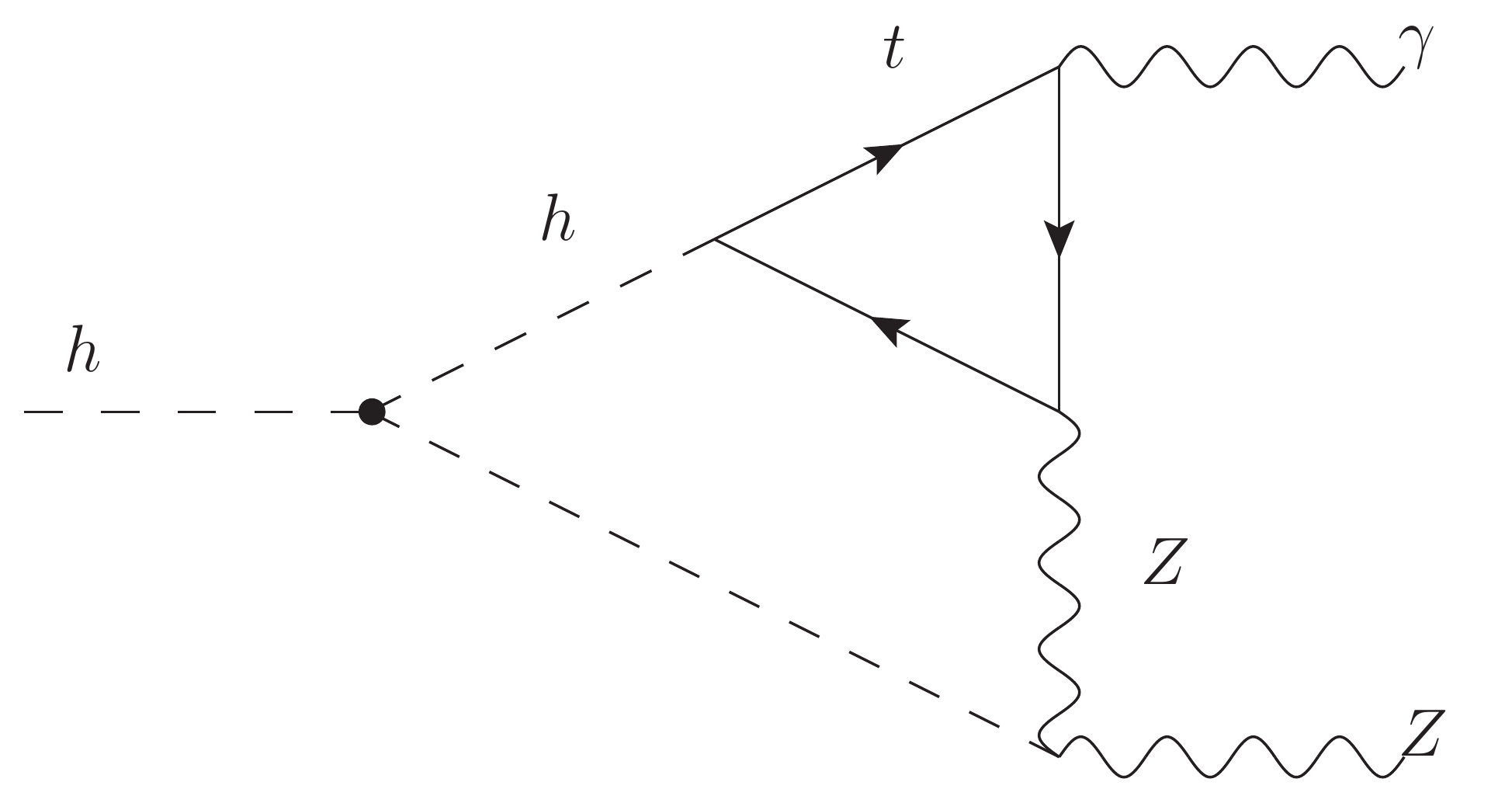}
	}
	\subfloat[]{
	\includegraphics[scale=0.20]{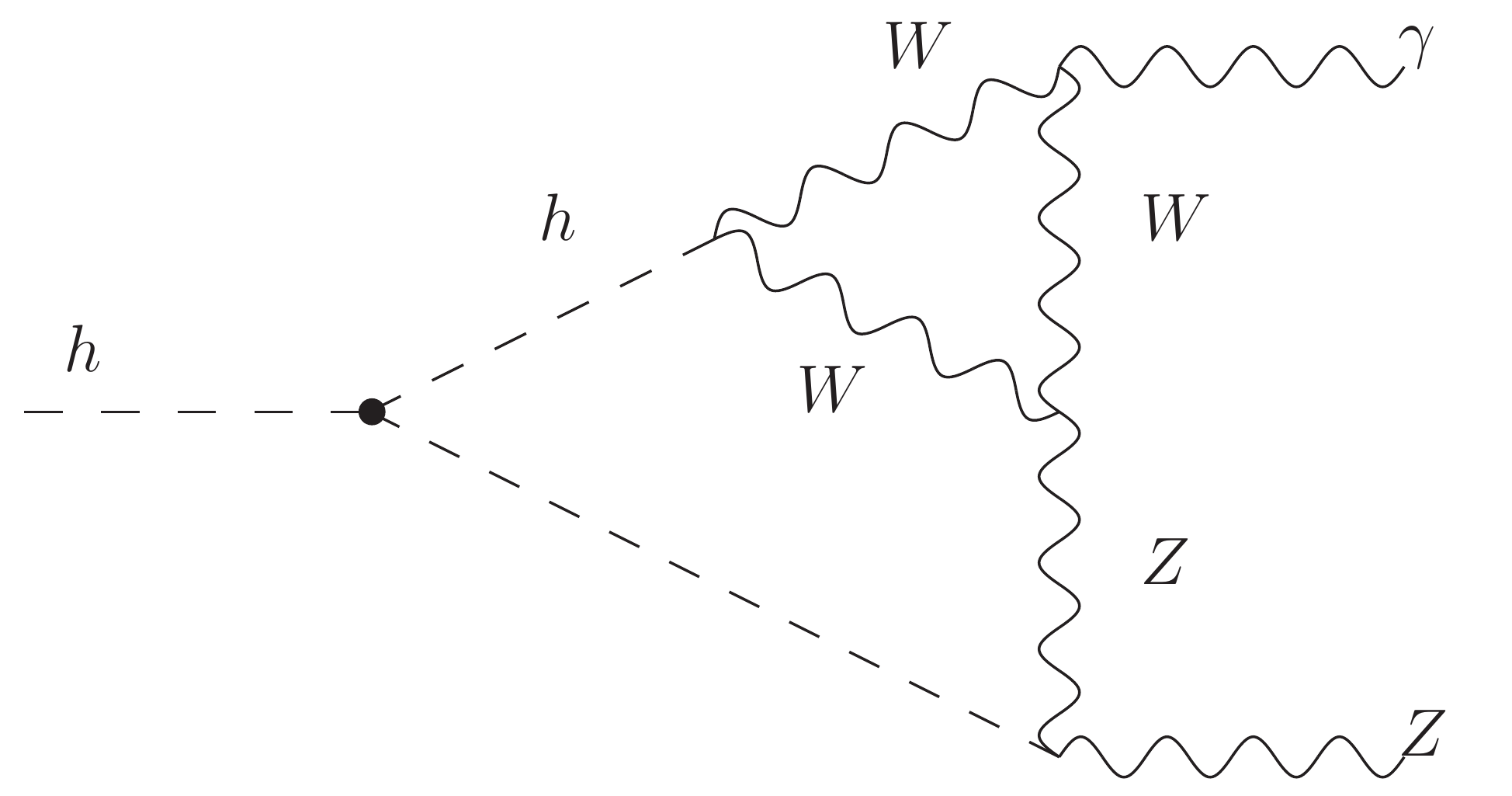} 
	}
        \caption{Examples of two-loop diagrams contributing to the
          $h\rightarrow \gamma Z$ amplitude:
          (a) diagram contributing to $\mathcal{F}_t$;
          (b) diagram contributing to $\mathcal{F}_W$. }  
\label{fig:1loop}
\end{figure}

The $C_1$ coefficient in eq.(\ref{corr}) can be defined as
\cite{Degrassi:2016wml}
\beq
C_1= \frac{\int d\Phi ~ 2  \text{Re}\Bigl(\mathcal{M}^{0 *}
  \mathcal{M}^1_{\lambda_3^{\text{SM}}}\Bigr)}{\int d\Phi ~|\mathcal{M}^0|^2}
\eeq
where the integration in $d \Phi$ is over the phase space of the final-state
particles, $\mathcal{M}^{0}$ is the Born amplitude and
$\mathcal{M}^1_{\lambda_3^{\text{SM}}}$ is the $\trilsm$-linearly-dependent
contribution in the the loop-corrected amplitude evaluated in the SM.
However, since in the $h \to \gamma Z$ case the phase-space integral is just
a multiplicative factor and both
amplitudes are purely real, $C_1$ can be more easily written as
\begin{equation}
C_1 = \frac{2 ~ \mathcal{F}^{\rm{NLO}}_{\text{1PI}}}{\mathcal{F}^{\rm{LO}}}.
\end{equation}
where $\mathcal{F}^{\rm{NLO}}_{\text{1PI}}$ represents the one-particle irreducible
(1PI) two-loop diagrams  containing an $h^3$ interaction.

In order to evaluate  the $C_1$ coefficient we
generated, in the UG,  the two-loop diagrams contributing to the
$h \to \gamma Z$  amplitude using the Mathematica package
\texttt{FeynArts}\cite{Hahn:2000kx}.
As in the one-loop case the diagrams can be assigned to the two
categories $\mathcal{F}_t, \, \mathcal{F}_W$, see fig.~~\ref{fig:1loop}.
The  diagrams were manipulated
using the package \texttt{FeynCalc}\cite{Mertig:1990an,Shtabovenko:2016sxi },
expanded in the external momenta and reduced to scalar integrals
using a private code.  After the reduction to scalar integrals we were
left with the evaluation of two-loop vacuum integrals that were computed
analytically using the results of Ref.\cite{Davydychev:1992mt}. 

The result for $C_1$ is automatically finite in the unitary gauge, i.e. no
renormalization is needed,  since the LO result does not depend
on the trilinear coupling.  As expected the fermionic and
bosonic contributions are separately  finite.

\section{$\tril$-dependent contribution in a $(\Phid \Phi)^n$ theory}
\label{sec:3}
This section is devoted to discuss how the result obtained in the
$\kappa$-framework working in the UG can be
recovered using a SM Lagrangian with a modified scalar potential of the form
\beq
V^{NP} = \sum_{n=1}^N c_{2n} ( \Phid \Phi )^n\,,  \qquad\qquad 
\Phi = \binom{\php}{\frac{1}{\sqrt{2}}( v+ \phu + i \phd )}\,,
\label{potential}
\eeq
working  in a renormalizable gauge that we choose for simplicity to be
the Feynman one (FG). In eq.(\ref{potential})
$N$ can be a finite integer or infinite, and in the latter case we
assume the series to be convergent, while    
the SM potential is recovered setting $N=2$ 
with $c_2 = - m^2$ and  $c_4 = \lambda$, where $-m^2$ is the Higgs mass term in
the SM Lagrangian in the unbroken  phase.

A simplified  discussion of the equivalence of the $\kappa$-framework
with a  $(\Phid \Phi)^n$ theory, when the modification of the trilinear
Higgs self-coupling appears at the two-loop level, was presented in 
Ref. \cite{Degrassi:2017ucl}. In that reference the renormalization of the
scalar potential in eq.(\ref{potential}) was not discussed in detail
because it was not needed. While  the calculation of $\Gamma(h \to \gamma Z)$
in the $\kappa$-framework is automatically finite,  the
one in  a $(\Phid \Phi)^n$ theory requires  to address the renormalization
of the scalar potential. It is then natural to try to devise a renormalization
procedure such that one obtains automatically the same result in the two
approaches.

The potential $V^{NP}$ up to quintic interactions can be written as
\beqn
V_{5\phi}^{NP} &=& v\, \tau\, \phu + \tau \left[\php\phm + \frac12 \phd^2 \right]
+ \frac12 \mh^2 \,\phu^2 + \left( \frac{\Mh^2}{2 v}  + v\,d \tril\right)\phu^3
+ \nonumber\\
&& \frac{\Mh^2}{ v} \,\phu \left[ \php\phm + \frac12 \phd^2  \right]
+ \frac{\Mh^2}{2\,v^2} \left[\php\phm + \frac12 \phd^2 \right]^2 +
 \nonumber\\
&&   \left( \frac{\Mh^2}{2\, v^2} + d \lambda_4\right)
\frac14 \phu^4 
+ \left( \frac{\Mh^2}{2 v^2}  + 3 \,d \tril \right)  \phu^2 
\left[ \php\phm + \frac12 \phd^2 \right] + \nonumber\\
&&  \frac1v \left\{d\lambda_5 \phu^5  
 + \left( -3 \,d \tril  + d \lambda_4 \right)
 \phu^3 \,\left[\php\phm + \frac12 \phd^2 \right] + \right. \nonumber\\
 && ~~~~~~\left.  
 3 \,d \tril\, \phu\,\left[\php\phm + \frac12 \phd^2 \right]^2
 \right\}~.
\label{PotqP}
\eeqn

The condition of the minimum of the potential reads
\beq
 \left.\frac{d \, V^{NP}}{d \, \phu} \right|_{\phu=0} = v\,\tau =
v\,\sum_{n=1}^N c_{2n}\,  n \left(\frac{v^2}{2}\right)^{n-1} =0 ~,
\label{minV}
\eeq
and
\beq
\mh^2 = 
v^2\,\sum_{n=1}^N c_{2 n}\,  n(n-1) \left(\frac{v^2}{2}\right)^{n-2} +
\sum_{n=1}^N c_{2 n}\,  n \left(\frac{v^2}{2}\right)^{n-1} \equiv
\Mh^2 + \tau
\label{Hmass}
\eeq
that enforcing the minimum condition implies $\mh^2 = \Mh^2$.
 The anomalous contributions in eq.(\ref{PotqP}) are:
\beqn
d \tril &=& 
\frac13 \sum_{n=3}^N c_{2n} \, n(n-1)(n-2) \left(\frac{v^2}{2}\right)^{n-2} \,,
\label{dc3}\\
d \lambda_4 & =& \frac23 \sum_{n=3}^N c_{2n} \, n^2(n-1)(n-2)  
\left(\frac{v^2}{2}\right)^{n-2}
\label{dc4}\\
d \lambda_5 &=& 
\frac1{30} \sum_{n=3}^N c_{2n} \, n(n-1)(n-2)(\frac32-4 \,n+ 2 n^2)
\left(\frac{v^2}{2}\right)^{n-2} 
\label{dc5}~.
\eeqn
Eqs.(\ref{dc3}--\ref{dc5}) give rise to  anomalous trilinear and
quadrilinear Higgs self-interactions as well as a quintic one. 
The $h^5$ interaction proportional to $d \lambda_5$ is not relevant for our
discussion while from the $h^3$ interaction we are going to identify
$\ktre = 1 + 2 v^2/\mh^2 \, d\tril$.

\begin{figure}[t]
\begin{center}
	\subfloat[]{
	\includegraphics[scale=0.27]{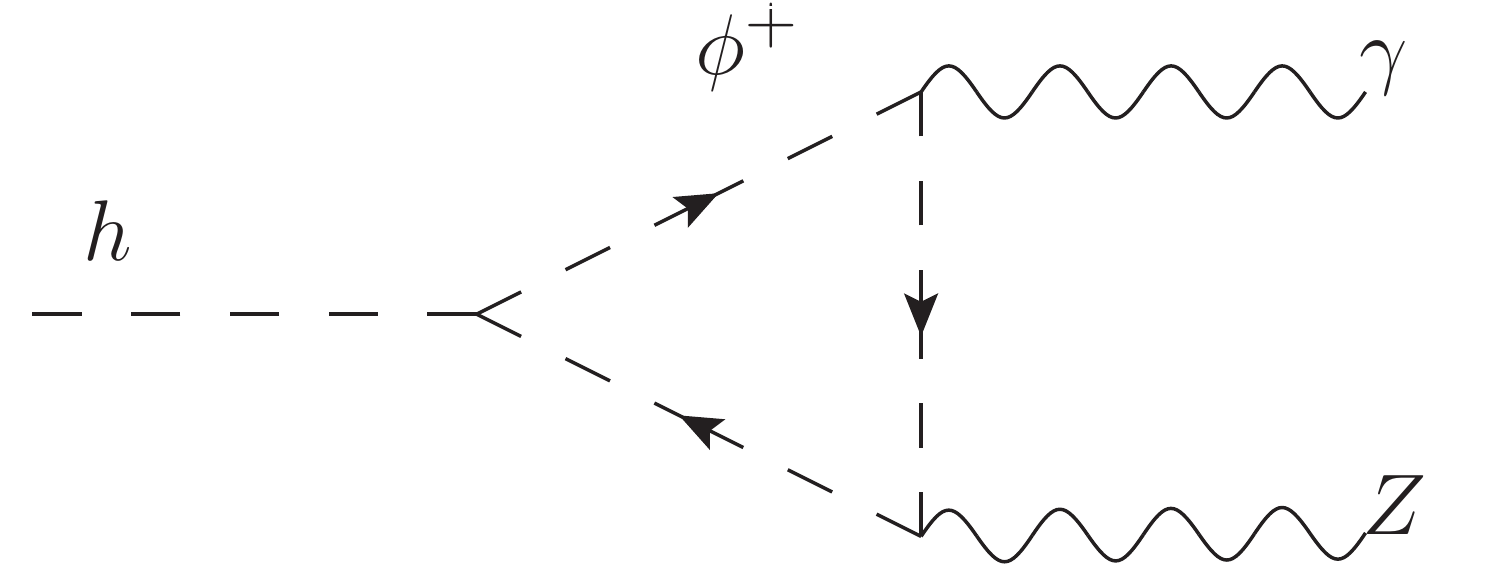}
	}
	\subfloat[]{
	\includegraphics[scale=0.27]{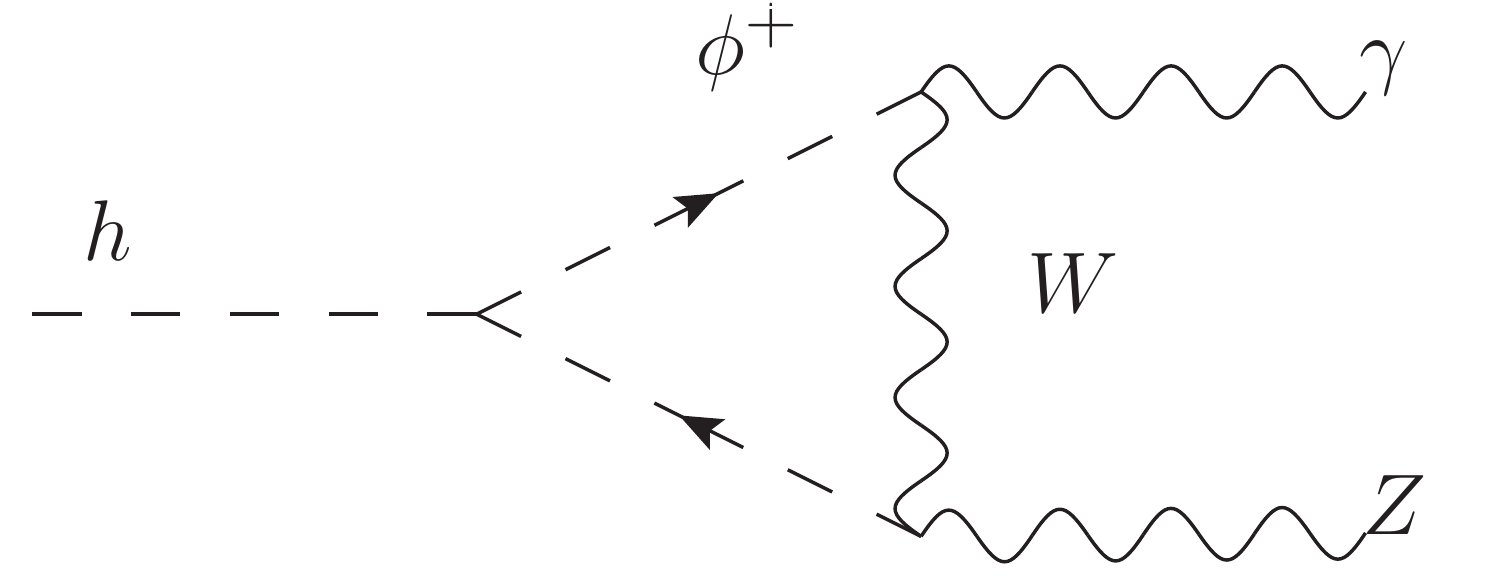} 
	}
	\subfloat[]{
	\includegraphics[scale=0.27]{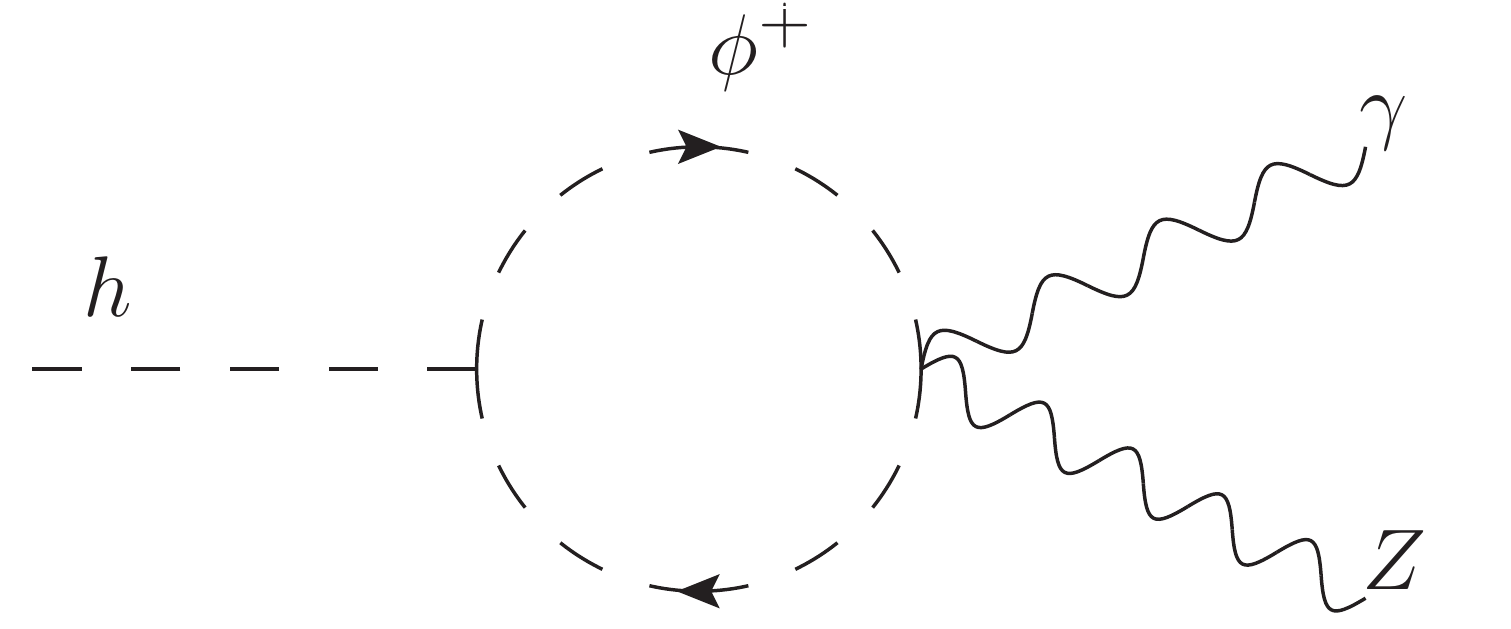} 
	}
\caption{ Examples of one-loop diagrams contributing in the FG to  the
  \mbox{$h \to \gamma Z$} process whose renormalization gives rise to a contribution
  proportional to an anomalous Higgs self-coupling.}
\label{fig:x}
\end{center}
\end{figure}

All the diagrams contributing to $h \to \gamma Z$ in  the UG at the
one-loop level contain only quantities (i.e. the gauge couplings,
the top and $W$ mass) whose one-loop renormalization is not affected by
Higgs self-interactions\footnote{The tadpole contribution is assumed to be fully
cancelled by the tadpole counterterm.}. Instead, the situation is different
in the FG
where, at one loop, there are diagrams, see fig.~\ref{fig:x}, that contain a
coupling proportional to $\Mh^2/v$ whose renormalization is affected by 
anomalous Higgs self-couplings. Then, the calculation in the FG
requires to set a renormalization procedure for $\Mh^2/v$ as well as
for the masses of the unphysical scalars.
  
\begin{figure}[t]
\begin{center}
	\subfloat[]{
	\includegraphics[scale=0.3]{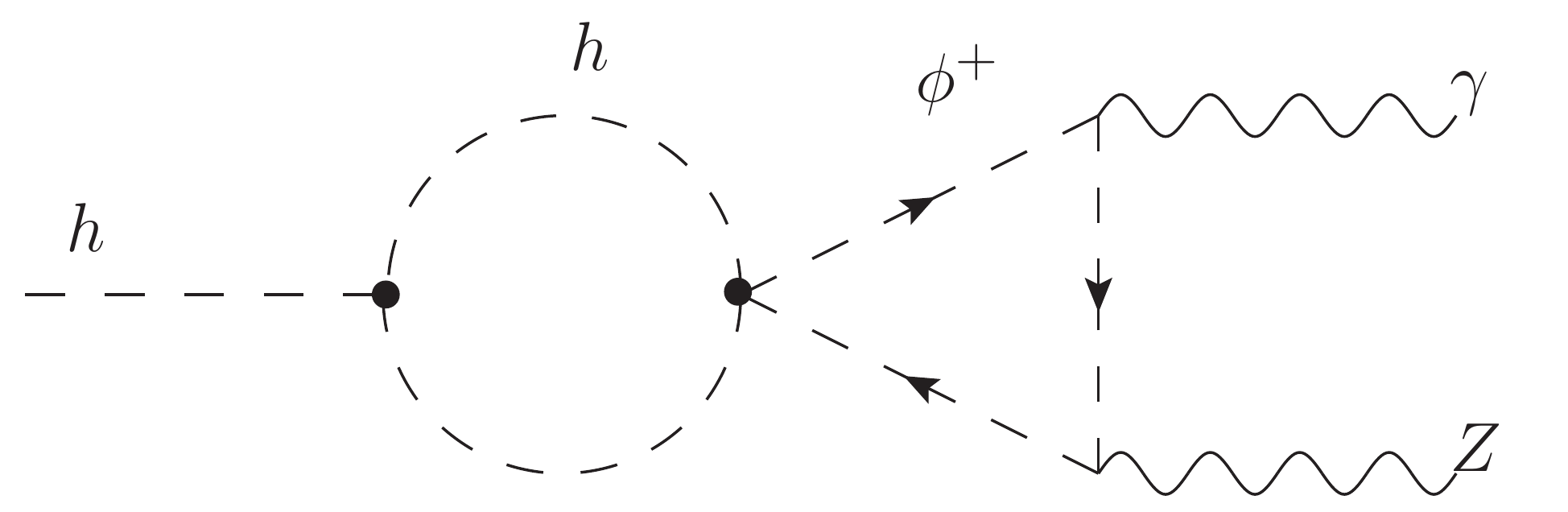}
	}
	\subfloat[]{
	\includegraphics[scale=0.3]{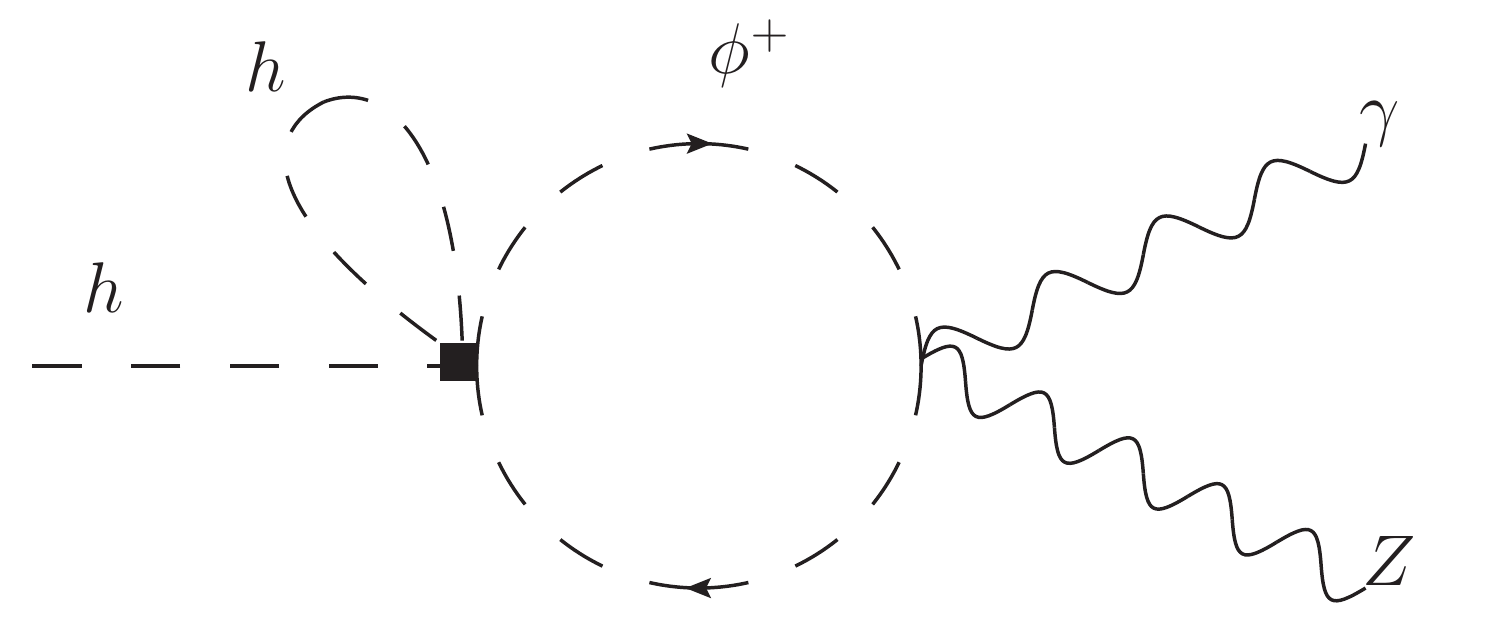} 
	}
\caption{ Examples of 1PI diagrams that contain contributions not present in
  the $\kappa$-framework  result. The dot (square) represents a
  coupling containing $d \tril$ 
  ($d \lambda_4$):  a) Contributions proportional to $(d \tril)^2$. b)
  Contributions proportional to $d \lambda_4$.}
\label{fig:y}
\end{center}
\end{figure}

Before discussing the renormalization of $V^{NP}$ we would like to notice
few things concerning the one-particle-irreducible (1PI) two-loop diagrams
in the FG. The 1PI diagrams contain terms proportional to $(d \tril)^2$,
i.e.~$(\ktre)^2$, see fig.~\ref{fig:y}(a), besides those from the
wave-function renormalization of the
external Higgs fields. Furthermore, there are 1PI diagrams proportional
to $d \lambda_4$ from the quintic $\phu^3 \php \phm$ coupling, see
fig.~\ref{fig:y}(b). Neither of these two kind of contributions are present in
the $\kappa$-framework result and we expect our renormalization procedure to
cancel exactly these contributions.

We are actually interested in defining the one-loop counterterms associated to
$\Mh^2,\: \tau$ and $v$. Following closely Ref.\cite{Sirlin:1985ux}
we assumed $V^{NP}$ to be written in terms of
bare quantities that are shifted according to: $c_{2n} \to c_{2n}^r -
\delta c_{2n},\: v \to v^r - \delta v$. As a consequence
\beq
V^{NP} = V^{NP}_r - \delta V^{NP}~,
\label{shiftV}
\eeq
where the renormalized potential up to quintic couplings has exactly
the same form given in eq.(\ref{PotqP}) but written in terms of renormalized
quantities, while the relevant terms in $ \delta V^{NP}$ that require to
be defined at one-loop are: $ \tau,\: \mh^2,\: \Mh^2,\: v$. 

To identify the  vacuum as the minimum of the 
radiatively corrected potential we set:
\beq
\delta( v\, \tau) = v\, \delta \tau =
v\,\sum_{n=1}^N \left(\frac{v^2}{2}\right)^{n-1} \left[ \delta c_{2n}\,  n 
+ c_{2n}\,  n (n-1)\,  2  \frac{\delta v}v \right]
  = - T
\label{eq:tadpole}
\eeq
where $i T$ is the sum of the tadpole diagrams with external leg extracted.
We identify $\mh^2$ in eq.(\ref{PotqP}) as the on-shell Higgs mass leading to
the condition (see eq.(\ref{Hmass}))
\beq 
\delta \mh^2 = \delta \Mh^2 + \delta \tau = {\rm Re}\, \Pi_{hh}(\mh^2)
\label{eq:mh}
\eeq
where  $ -i\, \Pi_{hh}(q^2)$ is the sum of all 1PI
self-energy diagrams and in eq.(\ref{eq:mh}) no tadpole contribution is present
because eq.(\ref{eq:tadpole}) is enforced.
Eqs.(\ref{eq:tadpole}-\ref{eq:mh}) imply
\beq
\delta \Mh^2 = {\rm Re}\, \Pi_{hh}(\mh^2) + \frac{T}v~.
\label{eq:Mh}
\eeq

The quantities ${\rm Re}\, \Pi_{hh}(\mh^2)$ and $T$ do contain terms
proportional to $d \tril$ and $d \lambda_4$ that will be relevant for our
discussion. The other quantity we are interested in is $v$, whose 
renormalization can be set, for example, either
from the muon-decay process or the $W$ mass and in both cases the corresponding
counterterm $\delta v$ does not contain (at one-loop) terms proportional
to an Higgs anomalous self-coupling and therefore will not enter in our
discussion. 

Eq.(\ref{eq:tadpole}) and eq.(\ref{eq:Mh}) are the two definitions  needed to
address the two-loop calculation in the FG of the effect
induced by an anomalous Higgs trilinear coupling on processes that are not
sensitive at the tree or one-loop level to an     $h^3$ interaction.

We find for the contribution proportional to $d \tril$ and $d \lambda_4$ in 
$T/v$ and ${\rm Re}\, \Pi_{hh}(\mh^2)$  in units
$\alpha/(4 \pi \sin^2 \theta_{W})\, \mh^2/( 8 \,\mw^2)$:

\beqn
      \frac{T^{d\lambda}}{v} &=&
   \left(1+ 2 \frac{v^2}{\mh^2} d \tril \right) \,3 \, \mh^2
  \left( \frac1{\epsilon} +1-    \ln \frac{\mh^2}{\mu^2} \right)
  \label{eq:tad} \\
    {\rm Re}\, \Pi^{d \lambda}_{hh}(\mh^2) & = &
     - \left( 1+ 2 \frac{v^2}{\mh^2} d \tril\right)^2\, 9\, \mh^2
   \left( \frac1{\epsilon} +2 -
   \frac{\pi}{\sqrt{3}} - \ln \frac{\mh^2}{\mu^2} \right)   \nn \\
      && 
  - \left(1+ 2 \frac{v^2}{\mh^2} d \lambda_4 \right) \,3 \, \mh^2
\left( \frac1{\epsilon} +1-    \ln \frac{\mh^2}{\mu^2} \right)  \nn \\
      &&
    -  \left( 1+ 6 \frac{v^2}{\mh^2} d \tril \right) \left[
        2\, \mw^2 \left(  \frac1{\epsilon} +1-\ln \frac{\mw^2}{\mu^2} \right)
        \right. + \nn \\
      && ~~~~~~~~~~~~~~~~~~~~~~~~~~~~~~~~~~ \left. 
        \mz^2 \left(  \frac1{\epsilon} +1-\ln \frac{\mz^2}{\mu^2} \right)
        \right] 
      \label{eq:Pi}
      \eeqn
where $\epsilon =(4 - nd)/2$, $nd$ being the dimension of the space-time,
    and   $\mu$ is the t-Hooft mass. 

From the inspection of eq.(\ref{eq:Pi}) one sees that the counterterm
associated to $\Mh^2$ (eq.(\ref{eq:Mh}))  contains terms proportional
to $(d\tril)^2$ and $d \lambda_4 $. It is easy to show that the renormalization
of the one-loop diagrams proportional to $\Mh^2$ (see fig.~\ref{fig:x}) cancels
exactly  the 1PI two-loop contribution  proportional to $(d \tril)^2$,
see fig.~\ref{fig:y}(a), as well as  the one  proportional
to $d \lambda_4$, see fig.~\ref{fig:y}(b), restoring in the FG the same
dependence on the anomalous coupling found in the $\kappa$-framework, i.e
linear in $\ktre$, apart the quadratic
contribution related to the wave function renormalization, and no dependence
on $\lambda_4$.

The renormalization conditions eq.(\ref{eq:tadpole}) specifies
also the part of the mass counterterms for the unphysical scalars that
is affected by anomalous Higgs self-couplings, i.e. $\delta \tau$.
A detailed discussion of the role of this counterterm in the cancellations
in the FG between 1PI and counterterm diagrams can be found in
Ref.\cite{Degrassi:2017ucl}.
    
\begin{figure}[t]
\begin{center}
\includegraphics[width=0.5\textwidth]{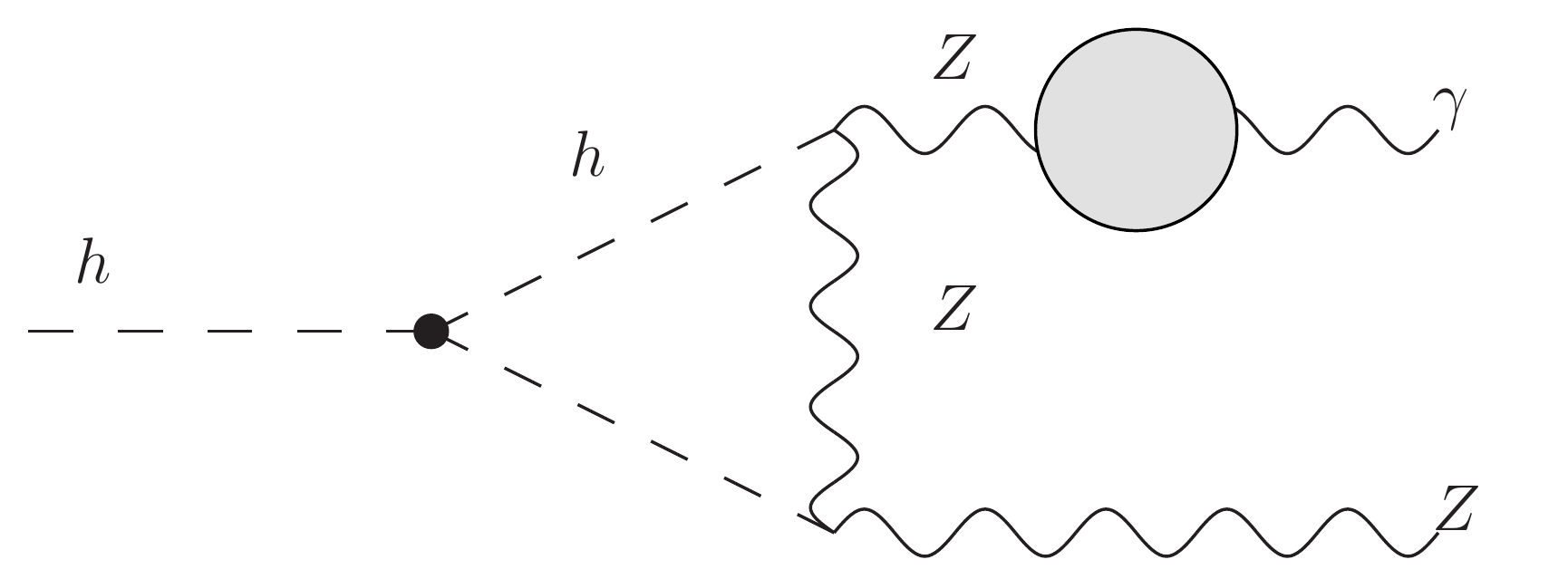}
\caption{ Two-loop reducible diagrams that  give rise to a contribution
  proportional to $d \tril$ in the FG. The meaning of the dot is as in
  fig.~\ref{fig:y}.}
\label{fig:z}
\end{center}
\end{figure}

To complete the calculation of the contribution of an anomalous trilinear
Higgs self-coupling in the FG one has to take into account also the
effect of  the reducible diagrams  shown in fig.~\ref{fig:z}.
In a renormalizable gauge diagrams that contain a $\gamma Z$ self-energy
evaluated at vanishing external momentum   also contribute to $\mathcal{F}_W$.
These diagrams do not contribute in the UG because in this gauge the
$\gamma Z$ self-energy evaluated at vanishing external momentum is zero.

We have computed the functions ${\cal F}_t$ and  ${\cal F}_W$ in the FG
and found agreement with the result in the UG within the order of our
approximation, as discussed  in the next section. We remark  that
with our choice of renormalization conditions the agreement is found in a
straightforward way because  both results are
expressed in terms of physical quantities.

As a final remark we notice that the potential in eq.(\ref{potential}) has
$N$ $c_{2n}$ coefficients, that are assumed to be bare quantities, while
we needed to impose only two renormalization conditions, eq.(\ref{eq:tadpole})
and eq.(\ref{eq:mh}). Furthermore, the modifications of the Higgs
self-couplings with respect to the SM values, $d \lambda_i, i=3..$, involve
only coefficients with $n \geq 3$ (see eqs.(\ref{dc3}-\ref{dc5})).
Then, on one side we have constructed a framework such that the  limit to the
SM case is straightforward.
On the other,  because we did not need  to specify any renormalization
condition on the modifications $d \lambda_i$, the  renormalization of the
latter is still free and can be specified via other processes.  

\section{Results}
\label{sec:4}
In this section we present the results for the $C_1$ coefficient. 
In table \ref{tab:uninum1} we give the numerical results for the first four
orders of the expansion of $\mathcal{F}_t^{\rm{NLO}}$ and $\mathcal{F}_W^{\rm{NLO}}$
up to and including terms of $\mathcal{O}(h_{4t}^3)$ and $\mathcal{O}(h_{4w}^3)$,
respectively.  The input parameters used are the same of
Ref.~\cite{Degrassi:2016wml}, or
\beq
\mw = 80.385,\: \mz = 91.1876,\:
\mh =125, \: \mt = 172.5
\eeq
where all the masses are in GeV.
\begin{table}[t]
\centering
\begin{tabular}{cccccc}
\hline
Fermionic	& $h_{4t}^0$		& $h_{4t}^1$ 	& $h_{4t}^2$
& $h_{4t}^3$ & Total \\
\hline
$\vphantom{\Bigl[}\mathcal{F}_t^{\rm{NLO}} (10^{-2})$ & 0.197 		& 0.086
  & 0.016 &  0.003 & 0.303 \\
$\mathcal{F}_t^{\rm{NLO}}$/$h_{4t}^0$ & 1.0 & 43.7\% & 8.1\% & 1.5\% &  \\
\hline
 Bosonic & $h_{4w}^0$ & $h_{4w}^1$ & $h_{4w}^2$ & $h_{4w}^3$ & Total \\
\hline
$\vphantom{\Bigl[}\mathcal{F}_W^{\rm{NLO}} (10^{-2})$ & -1.645		& -0.538
  & -0.129 & -0.043  & -2.355 \\
$\mathcal{F}_W^{\rm{NLO}}$/$h_{4w}^0$ & 1.0 & 32.7\%  & 7.9\%  & 2.6\% &  \\
\hline
\end{tabular}
\caption{Small-momentum expansion results for the two-loop
  $\lambda_3$-dependent contributions to the $h \to \gamma Z$
  amplitude.  The importance of each term
  in the expansion with respect to the first one is also shown.}
\label{tab:uninum1}
\end{table}

The table shows that  the expansions of both the fermionic and bosonic
contribution have a good convergence. In particular the last term in the
fermionic expansion, $\mathcal{O} (h_{4t}^3)$, contributes to the total fermionic
contribution at the level of 1\%, while in the bosonic case the last
term, $\mathcal{O} (h_{4w}^3)$, contributes to the total at the level of 2\%.
Given $\mathcal{F}^{\rm{LO}}=-5.29$ we find $C_1 = 0.72 \cdot 10^{-2}$, a value
larger than  the $C_1$ coefficient for the $h \to \gamma \gamma$ decay
($C_1^{\gamma\gamma}=0.49 \cdot 10^{-2}$) and close to the one of the
$h \to ZZ$ decay ($C_1^{ZZ}=0.83 \cdot 10^{-2}$) that is the decay mode
most sensitive to an anomalous trilinear coupling.

\begin{figure}[t]
\centering
\subfloat[]{
	\includegraphics[scale=0.5]{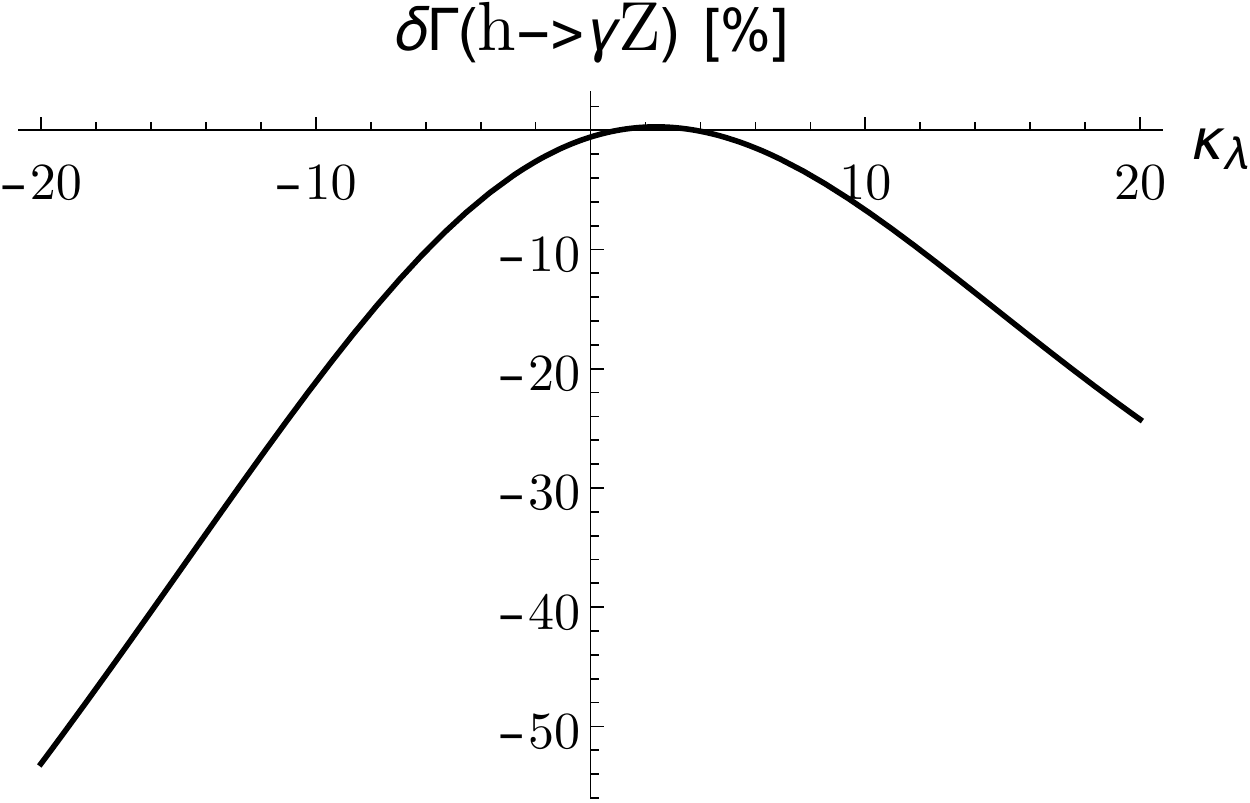}
	}
	\subfloat[]{
	\includegraphics[scale=0.5]{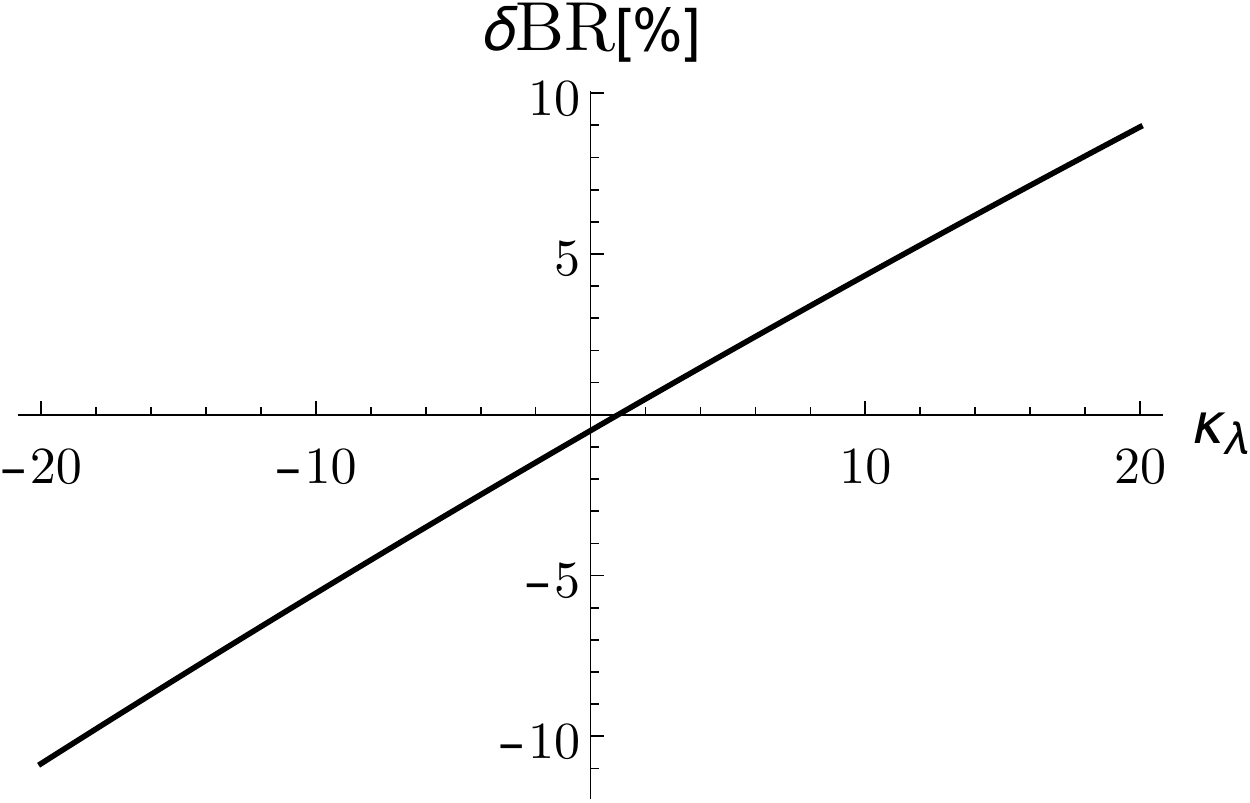}
	}
	\caption{Modification of the $h \to \gamma Z$ decay width (a) and
          branching ratio (b) due to an anomalous $\lambda_3$.}
\label{fig:deltagammabr}
\end{figure}

The effect of an anomalous $\lambda_3$ in the partial decay width
$\Gamma(h \to \gamma Z)$ and in the corresponding BR 
is presented in fig.~\ref{fig:deltagammabr} as a function of $\ktre$.
Similarly to the other decay modes of the Higgs boson \cite{Degrassi:2016wml},
the correction to the partial decay width can be substantial even for
$-10 \lesssim \ktre \lesssim 10$, while for the same range of $\ktre$ values
the correction to the BR is much smaller because in the BR the
universal quadratic dependence on $\ktre$ in eq.(\ref{corr}) cancels out.

Finally we want to comment on the numerical agreement between the calculation
using the $\kappa$-framework working in the UG and the one in
a $(\Phid \Phi)^n$ theory working in the FG.
While the results of the expansion for the fermionic contribution in
the UG and in FG are equal at the analytic level term by term, the same is not
true for the bosonic contribution. The reason is related to the fact that
the expansion in the external momenta is, in general, not the same in the two
gauges. Indeed, part of the kinematic dependence in the UG is transfer in a
FG to couplings as can be easily understood looking at the one-loop diagrams.
The diagrams in fig.~\ref{fig:x}, that appear in the FG, are proportional to
$\mh$ from the coupling $\phu \php\phm$, while in the UG  the only dependence
from $\mh$ at one-loop is of kinematical origin when the external momenta are
evaluated on-shell. In this situation we expect that the evaluation of
two-loop integrals via an  expansion in kinematical variables will not
give, at a fixed order, the same number in the UG and FG. However, if the
expansion is made correctly, we expect the numerical difference between the
two results to be of higher order with respect to the last known term. 

Defining the quantity
\begin{equation}
  \Delta \mathcal{F}_W =
  \Biggl|\frac{2 (\mathcal{F}_W^\text{UG}-\mathcal{F}_W^\text{FG})}{
    (\mathcal{F}_W^\text{UG}+\mathcal{F}_W^\text{FG})}\Biggr|
\end{equation}
as the relative difference between the results in the UG and the FG,
we find $\Delta \mathcal{F}_W =2.6\%$ for the total result, that is
well within the expected accuracy of an expanded result up to and
including $h_{4w}^3$ terms. In table \ref{tab:relativeUGFG}
we present $\Delta \mathcal{F}_W$ order by order in the expansion,
to show the  nice convergence pattern
between the numerical values in the UG and  in the FG.

\begin{table}[t]
\centering
\begin{tabular}{cccccc}
\hline
 & $h_{4w}^0$ & $h_{4w}^1$ & $h_{4w}^2$ & $h_{4w}^3$ & Total\\
\hline
$\vphantom{\Bigl[}\mathcal{F}_W^\text{UG} (10^{-2})$ & -1.645		&
  -0.538	& -0.129 & -0.043  & -2.355 \\
  $\vphantom{\Bigl[} \mathcal{F}_W^\text{FG} (10^{-2})$ & -1.181 & -0.727 &
    -0.280 & -0.113 & -2.301\\
$\Delta \mathcal{F}_W$& 32.8\% & 13.4\% & 5.5\% & 2.3\% & \\
\hline
\end{tabular}
  \caption{Relative importance of the difference between the UG and FG results
    for the bosonic contributions, for different orders of the expansion.}
\label{tab:relativeUGFG}
\end{table}

\section{Conclusions}
\label{sec:5}
In this work we have discussed the modifications in the partial decay
width $\Gamma(h \to \gamma Z)$ and in its BR induced by an anomalous
trilinear Higgs self-coupling. The two-loop computation has been
performed in the $\kappa$-framework and in a $(\Phid \Phi)^n$ theory.
In the latter case we had to address the renormalization of the scalar
potential. We showed that the conditions on the minimum of the
potential (eq.(\ref{eq:tadpole})) and on the Higgs mass
(eq.(\ref{eq:Mh})) are sufficient in order to obtain a finite result
in the $(\Phid \Phi)^n$ theory in agreement with the one  obtained in the
$\kappa$-framework within the order of our approximate calculation. It
should be remarked that the two above renormalization conditions do
not actually specify the Wilson coefficients of operators with
dimension larger than 4.

We have found that the sensitivity of the $h \to \gamma Z$ to an
anomalous trilinear Higgs self-coupling process is very similar to
that of the $h \to WW$ mode that is the second most sensitive mode
after the $h \to ZZ$ one.

The same renormalization framework employed in this work can be used to
discuss  the $h \to \gamma \gamma$ decay in a $(\Phid \Phi)^n$ theory or,
if only the terms $ n\leq 3$ are considered, in the SMEFT with only the $O_6$
operator. For this decay there are in the literature two results, not
in agreement, one obtained in the $\kappa$-framework via an expansion in the
external momenta
up to and including terms $h^3_{4t}$, $h^3_{4w}$ \cite{Degrassi:2016wml} and
one in the SMEFT where the diagrams were evaluated in the limit $\mw \to
\infty$ or equivalently $\mh \to 0$ \cite{Gorbahn:2016uoy}.

We proved that the result obtained using a $(\Phid \Phi)^n$ theory with the
renormalization conditions discussed in section \ref{sec:3} is identical,
at the analytic level term by term, with that obtained in the
$\kappa$-framework confirming the result of Ref.\cite{Degrassi:2016wml}.

It is not easy to pin down the source of discrepancy with the result in
Ref.\cite{Gorbahn:2016uoy}. We suspect that it  could be due
to a different way of taking into account  the renormalization of
$\Mh^2$  that  gives a contribution also in the limit $\mh \to 0$
taken in that work.

\section*{Acknowledgements}
The authors thanks F.~Maltoni for useful comments.
This work was partially supported by the Italian Ministry of Research (MIUR)
under grant PRIN 20172LNEEZ.
\bibliographystyle{JHEP}
\bibliography{DV}

\end{document}